\documentclass[preprint,superscriptaddress,amsmath,aip,pop,showkeys, floatfix]{revtex4}
\usepackage{hyperref}
\usepackage{natbib}
\usepackage{graphicx}
\usepackage{amsmath}
\usepackage{bm}

\newcommand*{\vek}[1]{ \bm{#1} }
\newcommand*{\hl}[2]{ #1 \cdot 10^{#2} }

\begin{document}

\title{Numerical Simulation of Current Sheet Formation in a Quasi-Separatrix Layer
using Adaptive Mesh Refinement}
\date{\today}

\author{Frederic Effenberger}
\affiliation{Theoretische Physik IV, Ruhr-Universit\"at Bochum, Germany}
\author{Kay Thust}
\affiliation{Theoretische Physik I, Ruhr-Universit\"at Bochum, Germany}
\author{Lukas Arnold}
\affiliation{Institute for Advanced Simulation, Forschungszentrum J\"ulich, Germany}
\author{Rainer Grauer}
\affiliation{Theoretische Physik I, Ruhr-Universit\"at Bochum, Germany}
\author{J\"urgen Dreher}
\email{juergen.dreher@rub.de}
\affiliation{Theoretische Physik I, Ruhr-Universit\"at Bochum, Germany}

\begin{abstract}
The formation of a thin current sheet in a magnetic
quasi-separatrix layer (QSL) is investigated by means of numerical
simulation using a simplified ideal, low-$\beta$, MHD model.
The initial configuration and
driving boundary conditions are relevant to phenomena observed in the
solar corona and were studied earlier by
\textbf{Aulanier et al., A\&A \underline{444}, 961 (2005)}.
In extension to that work, we use the technique of adaptive mesh
refinement (AMR) to significantly enhance the local spatial resolution of
the current sheet during its formation, which enables us to follow the evolution
into a later stage.
Our simulations are in good agreement with the results of Aulanier et
al. up to the calculated time in
%as the current sheet is properly resolved by the numerical grid in
that work.
In a later phase, we observe a basically unarrested collapse of the
sheet to length scales that are more than one order of magnitude
smaller than those reported earlier.
The current density attains correspondingly larger maximum values
within the sheet.
During this thinning process, which is finally limited by lack of
resolution even in the AMR studies, the current sheet moves upward, following
a global expansion of the magnetic structure during the quasi-static evolution.
The sheet is locally one-dimensional and the plasma flow in its
vicinity, when transformed into a co-moving frame, qualitatively
resembles a stagnation point flow.
In conclusion, our simulations support the idea that extremely high
current densities are generated in the vicinities of QSLs as a response
to external perturbations, with no sign of saturation.
\end{abstract}


\maketitle

\section{Introduction}

The spontaneous formation of thin current sheets is believed to play
an important role in the dynamics of astrophysical plasmas like the
solar corona and, in particular, for the onset of magnetic
reconnection.
In fact, impulsive events like solar flares and coronal mass ejections
are often associated with magnetic reconnection as a mechanism
that effectively releases magnetic energy, which in turn calls for
highly concentrated electrical currents to explain the breakdown of
ideal plasma behavior by means of e.g. micro-instabilities in the
non-collisional coronal plasma.

Recently, the study of magnetic field configurations as
candidates for reconnection has shifted from separatrix surfaces
towards quasi-separatrix layers (QSLs)
\cite{demoulin1996a,demoulin1996b,Milano1999}.
In contrast to genuine separatrix configurations, QSL fields do not necessarily
contain magnetic null points in 3D, which makes them relevant in many more situations
than strict separatrices. QSLs describe a magnetic field mapping
between two boundaries that is continuous, but changes rapidly in space.
This change has been quantified in terms of a flux tube ``squashing factor'' $Q$
\cite{titovHornig2002} and more recently generalized to remove boundary
projection effects \cite{titov2007}.
In accordance with the significance of QSLs for magnetic reconnection,
the problem of current sheet formation has been investigated.
Here, Titov et al. \cite{Titov2003} and Galsgaard et al. \cite{Galsgaard2003} 
studied the formation of current sheets in a straight hyperbolic
flux tube (HFT) configuration both analytically and numerically, 
and found exponential growth of the current density
as a response to shearing magnetic footpoint motion.
One major conclusion of that work was that a shear in the applied
boundary motion was important, if not essential, for the formation of
thin current sheets, and that the creation of a stagnation point flow
in the HFT was the key element in this process.  However, this
has been later questioned by Aulanier et al. \cite{aulanier2005}:
They argued that the initial squashing factor $Q$ in the simulations of
Galsgaard et al., probably together with additional symmetry
properties, was too small to account for highly concentrated currents, and
that the shear boundary motion and stagnation point flow 
in \cite{Galsgaard2003} served as
to dynamically create thin QSLs  only during the simulation.
This argument was underpinned in \cite{aulanier2005} through extensive
comparison between MHD simulations of less symmetric potential
magnetic field configurations which initially result from four
magnetic point sources submerged below a photospheric boundary. They
contain QSLs with squashing factors of up to $Q \approx 10^5$ and were
exposed to different boundary driver patterns, namely a shear and a
translational motion. Aulanier et al. observed the formation of thin, intense
current sheets, almost irrespective of the type of applied boundary
driving, which stresses the relevance of the initial field geometry
and QSL strength, rather than the boundary motion, for current sheet formation.
Later, that work has been extended by finite resistivity to simulate
magnetic reconnection at that thin current sheet \cite{Aulanier2006}
with its strong temporal change in magnetic connectivity.

A general limitation of these previous numerical studies was the lack
of reasonable numerical grid resolution at the sites of current sheet
formation \cite{Aulanier2006}: During the formation process, the currents get 
more and more concentrated locally so that the sheet scales quickly 
reach the numerical grid spacing.
This left open the question whether the current concentration would continue
to length scales small enough to account for the onset of micro-instabilities,
or if this process would get arrested at some larger length scale.
In order to address this question, but also get more insight into the
local dynamics in the immediate vicinities of the current sheets, we
have carried out numerical simulations in the spirit of \cite{aulanier2005},
using one of their initial configurations and boundary drivers.
However, we used the technique of adaptive mesh refinement (AMR),
implemented in the code {\sl racoon} \cite{drehergrauer}, in order to
obtain a much higher grid resolution in the current sheet vicinity
than was reported earlier.
This allowed us to obtain insight into the formation process that reaches
beyond the previous results.

In the next section, we will briefly describe the model that we employed in our
studies, while results that complement the work of \cite{aulanier2005} ~ are given in 
Sec. III, followed by a summary and conclusion in the last section.

\section{Basic equations and numerical model}

Our simulations are based on a reduced subset of the MHD equations,
appropriate for the quasi-static, low-$\beta$ regime, where $\beta$ is
the ratio of thermal to magnetic pressure. In normalized quantities,
it reads
\begin{subequations}
\begin{eqnarray}
  \label{eq:momentum}
  \partial_t \vek{v} &=& -\vek{v}\cdot\nabla\vek{v} + \frac{1}{\rho}\vek{j} \times \vek{B} + \nu\Delta\vek{v}\\
  \label{eq:induction}
  \partial_t \vek{B} &=& \nabla \times \left ( \vek{v} \times \vek{B} \right )
  + \eta\Delta\vek{B} - \nabla\Psi\\
  \label{eq:ampere}
  \vek{j} &=& \nabla\times\vek{B} \\
  \label{eq:cleaning}
  \partial_t\Psi &=& -c_h^2\nabla\cdot\vek{B} + c_l\Delta\Psi
\end{eqnarray}
\end{subequations}
Pressure terms are omitted from the momentum equation
(\ref{eq:momentum}) because of the low-$\beta$ approximation.  As we
are interested in a quasi-stationary evolution of the system, i.e. the
limit of high wave speeds, we replace the continuity equation by a
direct prescription of a relaxation mass density, namely $\rho :=
B^2$.  This approach results in an homogeneous Alfv\'en velocity
$c_A := |B| / \sqrt{\rho} = 1$ and fast communication of
unbalanced forces throughout the system, which shortens relaxation
times and has been successfully used in other studies \cite{lukasFlarelab}.
Constant kinematic viscosity
$\nu=\hl{5}{-4}$ and resistivity $\eta=\hl{5}{-6}$ are included only
to guarantee numerical stability on the grid scale and have little
effect on the overall evolution. 
The artificial scalar function $\Psi$ and its dynamic equation
(\ref{eq:cleaning}) serve as a convenient means to constrain any
finite $\nabla\cdot\vek{B}$, resulting from discretization errors, to
negligible values \cite{dedner2002}:
Combining $\partial_t \nabla\cdot(\ref{eq:induction})$
and $ \Delta (\ref{eq:cleaning})$  results in the
mixed equation
$$\partial_{tt} \nabla\cdot \vek{B} = c_h^2\Delta \nabla\cdot \vek{B} 
+ (\eta+c_l)\Delta  \partial_t \nabla\cdot \vek{B}
- c_l\eta \Delta^2  \nabla\cdot \vek{B}$$
for $\nabla\cdot \vek{B}$.
The crucial term on the right side is the first Laplacian, which describes
a hyperbolic transport of $\nabla\cdot \vek{B}$ with velocity $c_h$ and leads to
radiative distribution of  $\nabla\cdot \vek{B}$ throughout the computational domain, while it gets
dissipated by phase mixing and diffusion according to the other two terms.
Note that there is freedom in a specific choice for the $\Psi$-equation (\ref{eq:cleaning})
as it is of the order of the discretization error anyway.
For instance, while Dedner et al.~\cite{dedner2002} discuss the term $-\frac{c_h^2}{c_p^2}\Psi$
(see their Eq. (19) resp. (24e))
to arrive at a telegraph equation for $\Psi$ and $\nabla\cdot\vek{B}$
in the continuous case, we found this unnecessary, although possible, for our computations. The main
reason for this seems to be that in our case, the sources of
$\nabla\cdot\vek{B}$-errors are highly localized regions in space,
namely the regions of intense currents, so that the hyperbolic
transport is the dominating cleaning effect.
On the other hand, 
we added the term $c_l\Delta \Psi$ in Eq. (\ref{eq:cleaning}), which is not discussed in this specific form
in \cite{dedner2002}.
Its motivation is, however, not so much a change in the  $\nabla\cdot\vek{B}$-cleaning
property itself, but the observation that the purely hyperbolic choice, i.e. Eq.~(\ref{eq:cleaning}) with $c_l=0$,
tends to introduce odd-even-decoupling on the centered finite difference grid that we used.
This decoupling, which was not an issue in \cite{dedner2002} since 
they employed finite volume schemes in finite element discretisations,
could be healed satisfactorily through the additional Laplacian term
which couples $\Psi$-values on neighboring grid points with each other.
We finally found overall good $\nabla\cdot\vek{B}$-cleaning
properties when choosing the numerical parameter values to $c_l=\hl{5}{-4}$ and
$c_h^2 = \hl{5}{-2}$.

The equations are discretized in a domain of $(x, y, z) \in [-0.7,
  0.7] \times [-0.5, 0.5] \times [0, 0.5]$, where we identify the
plane $z=0$ with the solar photosphere. Integration is performed with
a strongly stable third-order Runge-Kutta scheme \cite{shuOsherRK},
spatial differentiation is realized
with standard second-order finite differences.  In order to resolve
the expected small-scale features, we employed the block-structured
adaptive mesh refinement (AMR) framework {\sl racoon}
\cite{drehergrauer}. Here, we used the norm of the magnetic field gradient,
$\nabla\vek{B}$, to derive a local length scale for $\vek{B}$ that 
serves as an indicator criterion for local mesh refinement.
Effective local grid resolutions obtained with this technique were
$4096^3$ in the present work.

Initial conditions are adapted from \cite{aulanier2005}, where we address the
magnetic field configuration resulting from two pairs of opposite
polarity photospheric flux patches whose respective connecting axes
intersect at an angle of 150 degrees. Specifically, the initial
magnetic field stems from 4 virtual magnetic point sources, indexed by $i$,
below the photosphere
\begin{equation}
  \label{eq:bInit}
\vek{B}(\vek{r}) =  \sum_{i=1}^4 
F_i\frac{\vek{r}-\vek{r}_i}{|\vek{r}-\vek{r}_i|^{3}}
\end{equation}
with respective source strengths $F_1 = -F_2 = 1$ and $F_3 = -F_4 = 0.4$ and
locations $\vek{r}_{1,2} = (\pm \frac{1}{2}, 0, -\frac{1}{5})$ and 
$\vek{r}_{3,4} =(\mp \frac{\sqrt{3}}{20}, \pm \frac{1}{20}, -\frac{1}{10})$, respectively.  This field geometry
is known to contain two symmetric dome-shaped QSLs with squashing
factors of $Q\approx 10^5$, intersecting in a hyperbolic flux tube 
(see \cite{aulanier2005} for details).

In the course of the simulation, the field is exposed to a horizontal
photospheric vortex flow around the magnetic source $i=3$ in Eq. (\ref{eq:bInit}),
realized by prescribing the boundary
condition for $\vek{v}$ at $z=0$.  The maximum flow velocity is
$\max(|\vek{v}_{Bnd.}|) \approx \hl{2}{-2}$
and it gets ramped up in time according to
$\propto \frac{1}{2} \left[ \tanh(\frac{5}{2}(t-1) ) + 1 \right]$.

The detailed treatment of the lower boundary is as follows:
As all quantities are discretized as cell-centered variables,  boundary conditions have to provide
values for $\vek{v}$,  $\vek{B}$  and $\Psi$ at $z_{1/2} := \Delta_z/2$ in a way that is consistent
with the evolution equations and the overall second order accuracy in the grid spacing $\Delta_z$.
Denoting the boundary values of individual quantities at $z=0$ with a hat, the 
velocity $\hat{\vek{v}} = \vek{v}_{Bnd.}$ is explicitly given from the prescribed horizontal vortex flow, and in particular,
$\hat{v}_z = 0$.
All components of $\vek{v}$ at $z_{1/2}$ can now simply be interpolated in $z$-direction with second order accuracy between $z=0$ and
the interior grid points above $z_{1/2}$.
Using this direct forcing, there is no need to compute $\vek{j}$ or the Lorentz force at $z_{1/2}$ at all, hence we do not need
the horizontal components of $\hat{\vek{B}}$ at this stage.
Now, to integrate $\vek{B}$, we note  that the convective part of the $B_z$-equation becomes autonomous in the boundary plane,
involving no other undetermined quantities nor any derivatives normal to the boundary:
Writing the convective electric field as $\vek{\mathcal{E}} := -\vek{v}\times\vek{B}$ gives
$\hat{\mathcal{E}}_x = -\hat{v}_y \hat{B}_z$ and $\hat{\mathcal{E}}_y = \hat{v}_x \hat{B}_z$ so that
$\partial_t \hat{B}_z = -\nabla_h \cdot ( \vek{\hat{v}}_h \hat{B}_z )$ where the index $h$ indicates horizontal vector projections, 
e.g.~the prescribed $\vek{\hat{v}}_h = (\hat{v}_x, \hat{v}_y)$, and 
$\nabla_h := (\partial_x, \partial_y)$.
In other words, when ignoring the numerically motivated resistive diffusion and $\nabla\Psi$-terms, we are able to integrate the ``proper''
$\hat{B}_z$ completely for its own, which in turn allows us again to interpolate $B_z$ at $z_{1/2}$ up to  $O(\Delta_z^2)$, similar to $\vek{v}$
above.
Now, this ``proper'' $\hat{B}_z$-equation  also determines $\vek{\hat{\mathcal{E}}}_h$, which allows to update $\vek{B}_h$ at $z_{1/2}$.
At this point, the magnetic diffusion and $\nabla\cdot\vek{B}$-cleaning terms are taken into account as usual, where the former requires
an extrapolation of $\vek{B}_h$ across $z=0$.
Finally, the right side of Eq. (\ref{eq:cleaning}) is easily evaluated at $z_{1/2}$, using $\hat{B}_z$ from above and the Dirichlet condition
$\hat{\Psi}=0$.

While the upper and lateral boundaries can in principle be handled in the same fashion, using the no-slip and no-penetration
condition $\vek{v}=0$, we used a simpler approach there: Setting $\vek{v}$  to zero ahead of the boundaries, keeping
the tangential components of $\vek{B}$ fixed in ghost cells and
applying solenoidal and homogeneous Dirichlet conditions to the normal components of $\vek{B}$ and $\Psi$, respectively,
proved to be sufficient for those passive boundaries. Note, in particular, that no artificial Lorentz forces act there either.

The boundary treatment described above presumes a homogeneous numerical grid. It has been applied in the same spirit
to the AMR simulations that we present here, where additional complications occur at the interfaces between neighboring grid blocks
of different mesh resolution that abut the physical boundaries.
Without going into details, we only mention that additional
coarse-fine and fine-coarse interpolations are needed for the
computation of $\hat{B}_z$ and $\hat{\vek{\mathcal{E}}}$ at those
junctions, but that they do not pose any fundamental new challenge
apart from the programming complexity.

\section{Results}
\begin{figure}
\includegraphics[width=\hsize]{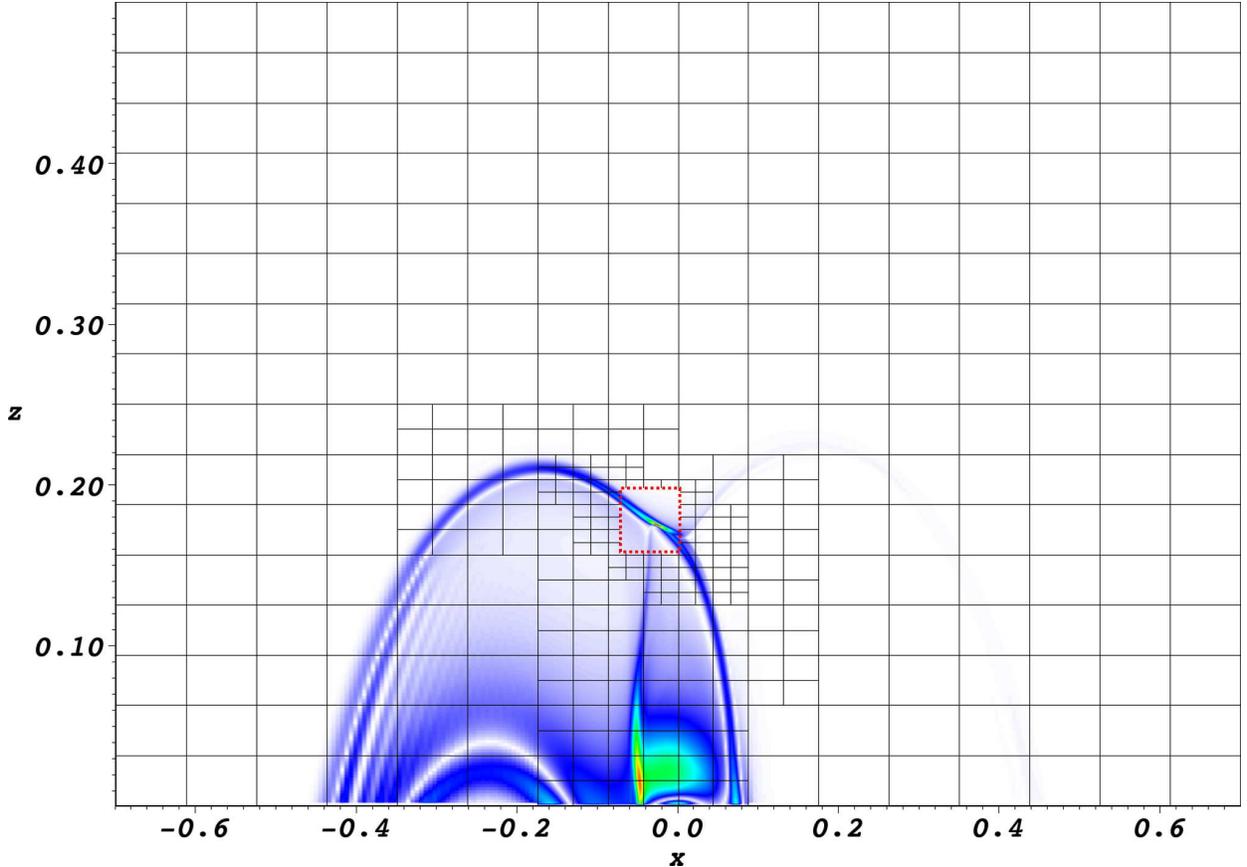}
\caption{
\label{figs:j_and_blocks}
Color coded $|\vek{j}|$ in the plane $y=0$ at $t=5.0$. Extended current systems match
those reported by \cite{aulanier2005}. The thin current sheet of interest formed inside
of the red rectangle and is hardly visible on these scales.
Maximum values of $|\vek{j}|$ are $ \approx 400$ near the photospheric boundary, 
and $\approx 1000$ in the marked region.
Block lines indicate the layout of the recursively refined
grid blocks, each containing $16^3$ cells. The two finest
block levels are omitted for clarity.
}
\end{figure}

After applying the photospheric boundary driving in $\vek{v}$,
dynamic shear modes travel into the domain and mix there.
On the scale of a few Alfv\'en transit times, the magnitude of the
current density grows significantly and a quasi-stationary
current system as shown in Fig.~\ref{figs:j_and_blocks} builds up.
It consists partly of relatively weak currents which are distributed on a large
scale in a dome-like structure that is pre-determined by magnetic
field lines connecting the driver region with the
opposite polarity regions of the photosphere.
On top of this, a highly localized thin current sheet can be identified 
in the vicinity of $(x, z) \approx (0, 0.18)$ in Fig.~\ref{figs:j_and_blocks}.
This thin current sheet actually lies inside the pre-existing QSL of
the initial magnetic field.
We interpret the striation patterns at $x\approx -0.4$ in
Fig.~\ref{figs:j_and_blocks} as signatures of MHD waves on field lines which connect the
strong photospheric field region with the current sheet during the
evolution.

These results are basically in good agreement with those published earlier
by Aulanier et al. \cite{aulanier2005}.
However, the sheet thickness at the stage shown in Fig.~\ref{figs:j_and_blocks}
is already on the numerical grid scale of \cite{aulanier2005}, so that their studies
were unable to investigate into the further evolution or features on smaller scales.
\begin{figure}
  \includegraphics[width=\hsize]{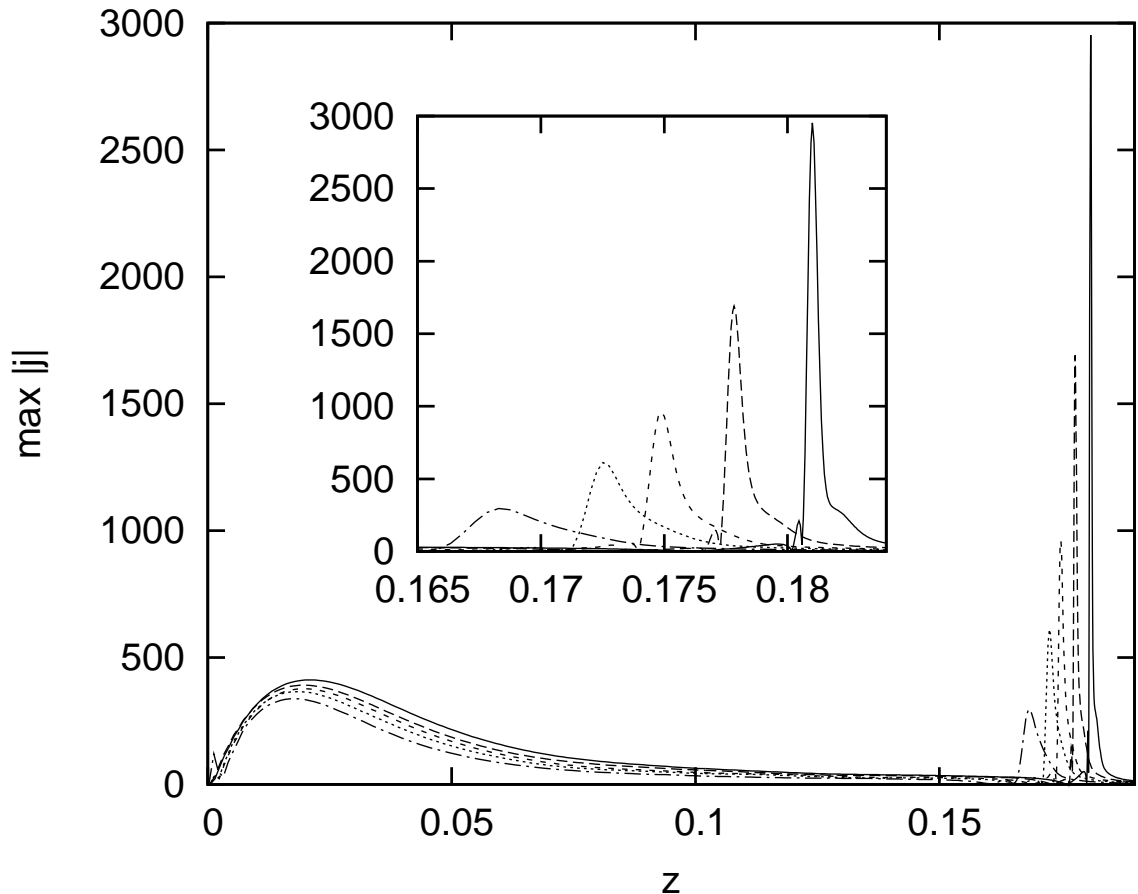}
  \caption{
    \label{figs:zProfiles}
    Plot of $|\vek{j}|$ vs. $z$ at $x=y=0$ taken at times $t=4.2,
    4.7, 5.0, 5.4, 5.8$.  The inset expands the relevant
    current sheet height range.  Maximum values of $|\vek{j}|$ are $\approx
    300, 600, 960, 1700$ and $2900$, respectively, increasing monotonically in time.
    The current sheet moves upwards and thins with
    respective FWHM values of $\approx 4.2\cdot 10^{-3}, 1.9\cdot
    10^{-3}, 1.2\cdot 10^{-3}, 7.2\cdot 10^{-4}$ and $4.8\cdot
    10^{-4}$.  }
\end{figure}
\begin{figure}
  \includegraphics[width=\hsize]{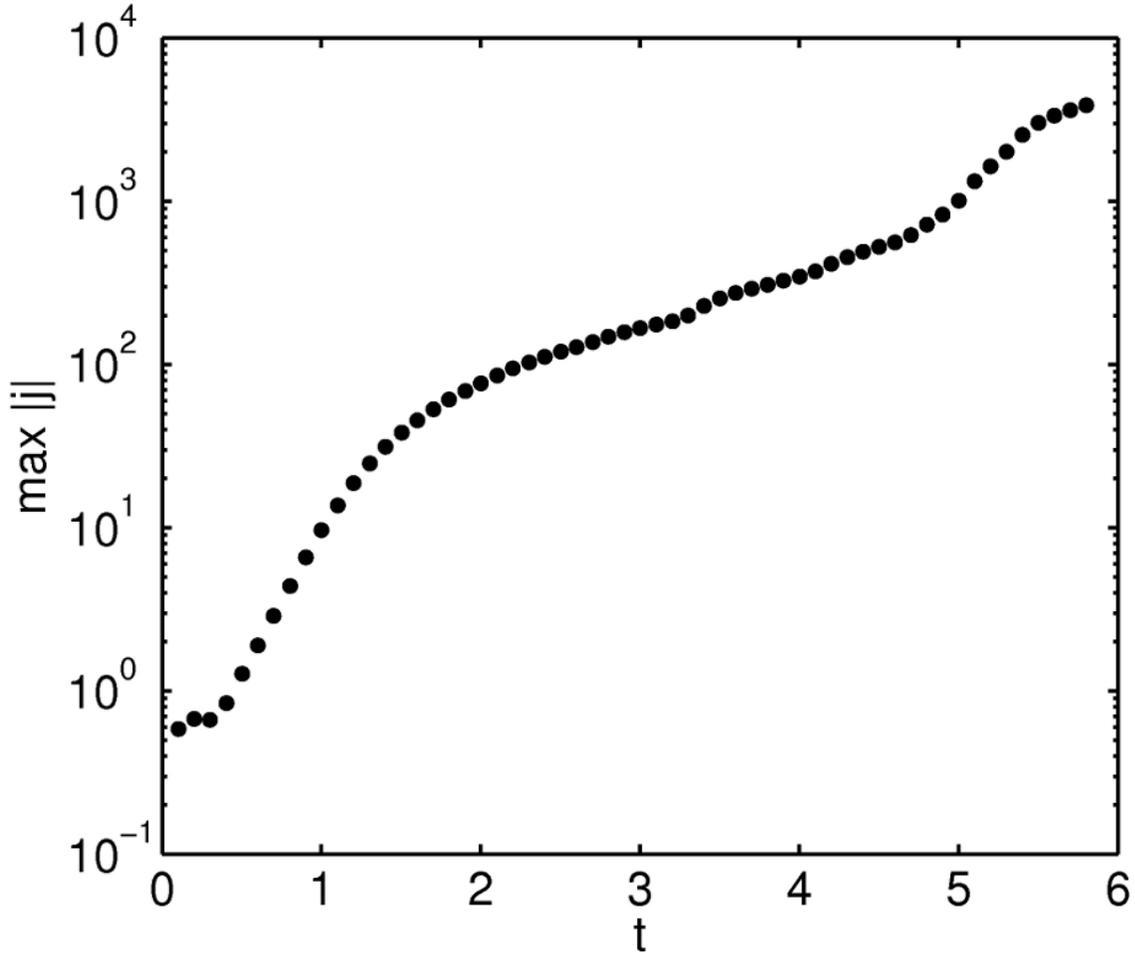}
  \caption{
    \label{figs:jmax_vs_t}
    Growth of the maximum of $|\vek{j}|$, taken over the entire domain, against time.
  }
\end{figure}

The temporal evolution of the sheet is displayed in Fig.~\ref{figs:zProfiles}
which shows the vertical profiles of $|\vek{j}|$ at $x=y=0$ for
four different times.
It is evident that the sheet thickness decreases with time while the value of
maximum current density increases accordingly to reach values of $\approx \hl{3}{3}$
in the latest stage.
At the same time, the sheet moves upwards, as indicated by the inset graphs.
In fact, we find that the entire magnetic structure expands gradually
as a response to the boundary driving, and the current sheet as a substructure
is embedded in this motion.
An other detail that emerges from Fig.~\ref{figs:zProfiles} is the
fact that up to $t\approx 4.4$, the most intense currents are not yet
found in the thin current sheet itself, but in the large-scale system at
$z\approx 0.02$, i.e. close to the photospheric driver (see
also Fig.~\ref{figs:j_and_blocks}).
It is only at later times, that the thin sheet dominates in the current intensity.

This phenomenon is also visible in the temporal evolution of 
$\max(|\vek{j}|)$, which is plotted logarithmically against time in Fig.~\ref{figs:jmax_vs_t}:
The early phase, with rather fast growth of $\max(|\vek{j}|)$,
corresponds to the ramp-up of the boundary driver, which
essentially reaches its maximum magnitude around $t\approx 1.7$.
This is followed by a slower growth rate of the current maximum up to
$t\approx 4.7$. During this stage, the maximum values stem from the
extended currents close to the photosphere (compare with Fig.~\ref{figs:zProfiles}), 
which eventually get overtaken by the faster growth of the thin embedded current sheet.
Further intensification continues, with amplification of
$\max(|\vek{j}|)$ by roughly a factor of 5, until the
growth slows down at $t\approx 5.5$.
At this time, the sheet thickness is only a few times the
numerical grid scale and thereby poorly resolved with artificial
diffusion effects becoming competitive.

\begin{figure*}
  \centering
  \includegraphics[width=\hsize]{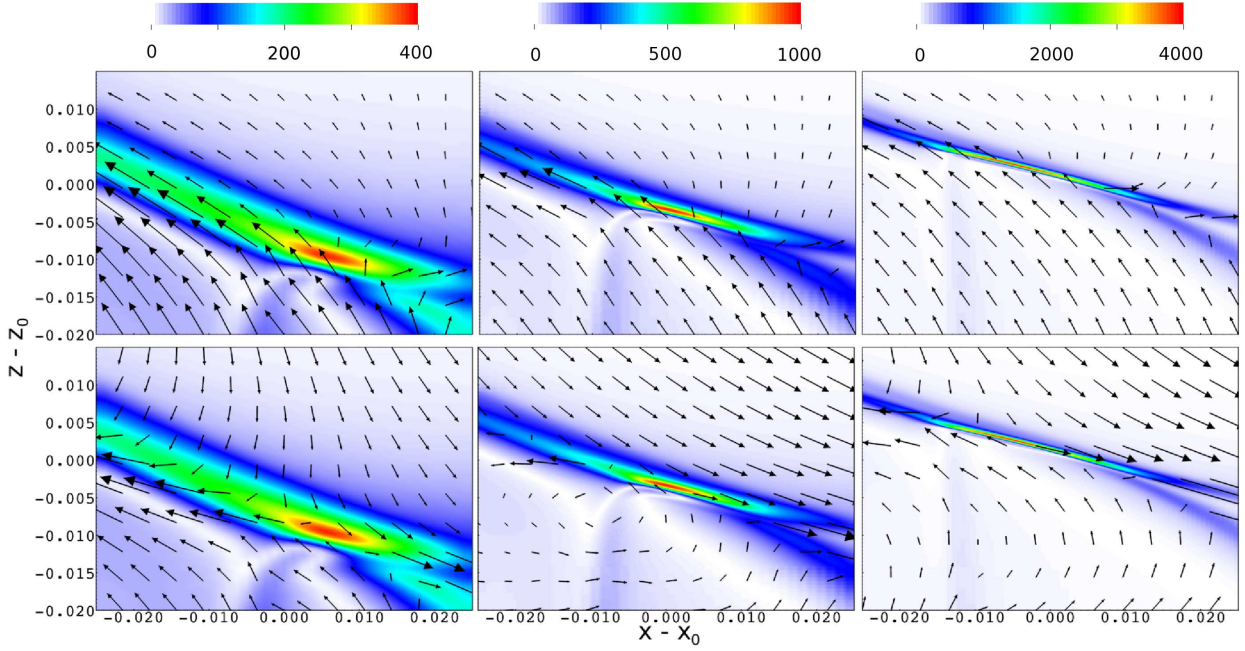}
  \caption{
    \label{figs:velocity}
    Color coded $|\vek{j}|$ and velocity components $(v_x, v_z)$ as arrows in the plane $y=0$
    at times $t=4.2, 5.0$ and  $5.8$ (left to right).
    Arrows in the upper row show the plasma velocity in the fixed reference frame,
    while $\vek{v}$ has been transformed into the co-moving frames of the current sheet
    for the lower figures. The transformation velocities are
    $(V_x, V_z) = (-3, 6) \cdot 10^{-3},  (-10, 8) \cdot 10^{-3}$ and $(-7, 7) \cdot 10^{-3}$,
    respectively with the transformed $|(v_x, v_z)|$ attaining maximum values of 
    $1.6\cdot 10^{-2}, 2.3\cdot 10^{-2}$ and $3.0\cdot 10^{-2}$ (left to right).
    Note also that the $x$- and $z$-coordinates on the axes are relative to the point
    $(x_0, z_0) = (-2.5, 18)\cdot 10^{-2}$.
    }
\end{figure*}
Fig.~\ref{figs:velocity} shows details of the sheet in the cut plane $y=0$ for three different times.
Again, the overall upward motion and the thinning and intensification of the sheet
are well visible.
In addition, we have plotted the plasma velocity as arrows, projected into the displayed
plane, to give an impression of the flow in the vicinity of the current sheet.
While the upper row shows the velocity relative to the fixed simulation frame,
the plots in the lower row show the flow transformed to a frame which is co-moving with the
expansion velocity of the structure.
To this end, we identified the locations of maximum current density
in the plane $y=0$ from plots of successive data output sets, and computed
a pattern velocity from their difference. This velocity was then subtracted
from the plasma flow velocity in the lower plots of Fig. 4.
There have been controversial discussions as to whether the current sheet formation 
at the hyperbolic flux tube embedded in the QSL is related to a stagnation-point
flow \cite{Galsgaard2003,aulanier2005}.
In particular, Aulanier et al. claim that no stagnation point exists at the 
intense current sheet.
This is certainly confirmed by our simulation, however we remark that the 
focus on a strict definition of stagnation point, i.e. $\vek{v}=0$, maybe somewhat
misleading because i) the velocity is sensitive to the chosen frame of reference,
and ii) the component along the current sheet should be discarded from these considerations
anyway, because it largely decouples from the mechanism of magnetic shearing discussed in
\cite{Titov2003} and \cite{Galsgaard2003}.
\begin{figure}
  \includegraphics[width=\hsize]{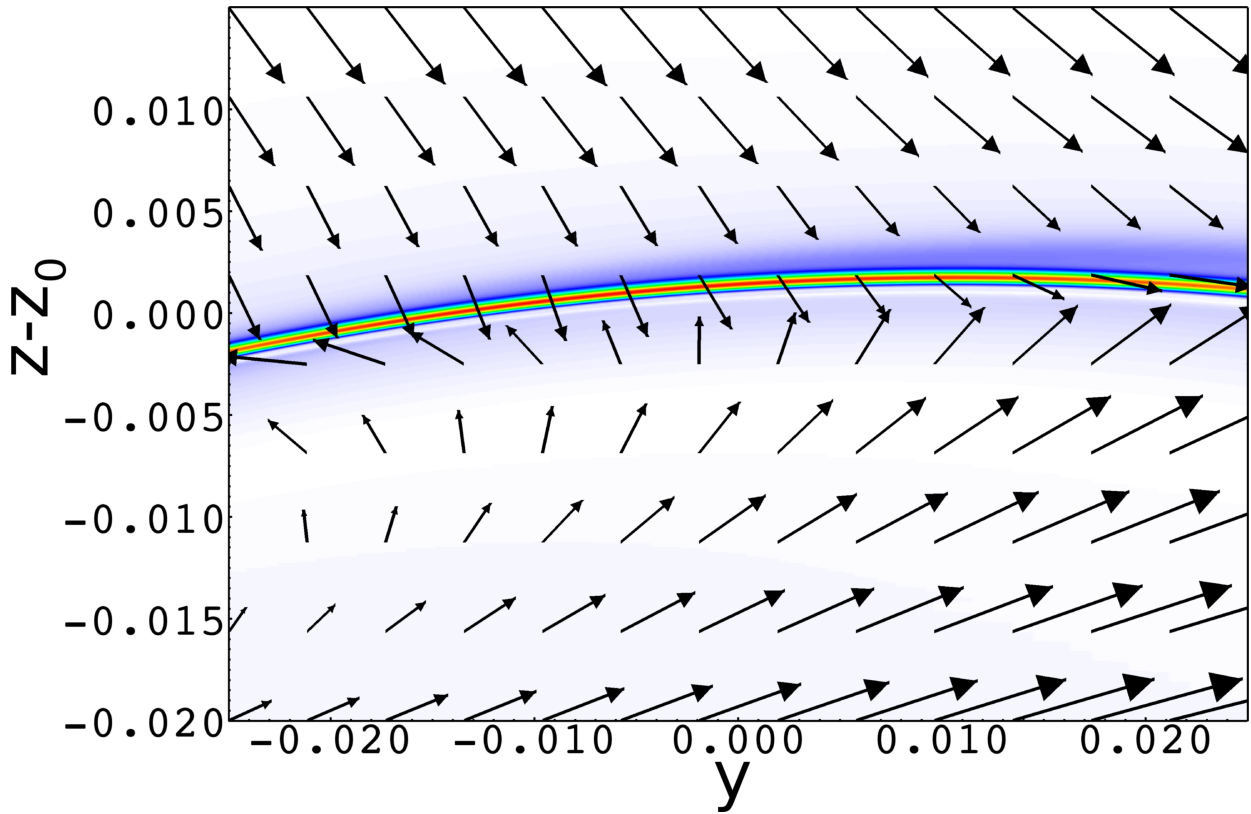}
  \caption{
  \label{figs:sliceX}
    Color coded $|\vek{j}|$ and velocity components $(v_y, v_z)$ at 
    $x=-2.5\cdot 10^{-2}$ and $t=5.8$, corresponding to the bottom right
    plot in Fig.~\ref{figs:velocity}. Here, $\vek{v}$ has been transformed into
    the co-moving frame $(V_y, V_z) = (-6, 7) \cdot 10^{-3}$ and 
    $\max(|\vek{v}|) \approx 1.1  \cdot 10^{-2}$ in that frame.
    Coordinates are relative to $(y_0, z_0) = (0, 0.18)$.
  }
\end{figure}
The flow pattern projected into the $y$-$z$-plane is shown in Fig.~\ref{figs:sliceX},
again transformed into a frame that moves upward with the current sheet and, in addition,
results in $v_y=0$ in the current sheet center.
This figure also demonstrates that the current sheet is indeed elongated in the $y$-direction.
Hence, at least in the latest stage displayed in Figs.~\ref{figs:velocity} and \ref{figs:sliceX},
it can be treated as a quasi-one-dimensional sheet.
\begin{figure}
  \includegraphics[width=\hsize]{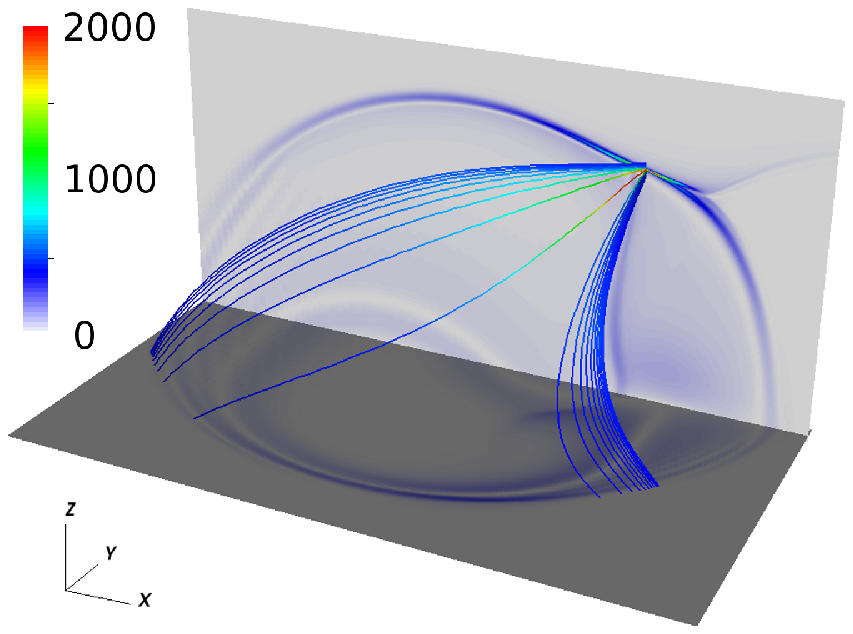}
  \caption{Color coded $\alpha = |\vek{j}|/|\vek{B}|$ in
  \label{figs:alphaNConst}
    the planes $y=0$ and $z=0$ at $t=5.8$.
    The maximum value $\alpha_m\approx \hl{2}{3}$ is attained in the
    current sheet center (red).
    The magnetic field lines, starting equidistant from
    $(-0.03, 0, 0.175)$ to $(-0.03, 0, 0.185)$, are also color coded
    with $\alpha$ and show that $\vek{B}\cdot\nabla\alpha\ne 0$,
    i.e. the magnetic field deviates significantly from a force-free
    field.
  }
\end{figure}
Finally, we remark that the assumption of a quasi-stationary evolution loses its
validity in the late stage of the sheet evolution: Obviously, the collapse becomes
a localized, dynamic process associated with significant magnetic forces.
This can be seen from the field line plot shown in Fig.~\ref{figs:alphaNConst},
where magnetic field lines connecting the thin current sheet with the photosphere have been
colored with the quantity $\alpha := |\vek{j}|/|\vek{B}|$. For a force-free
field, $\vek{j}\times\vek{B} = 0$, the value of $\alpha$ is constant along magnetic field lines.
This condition is obviously not met in the QSL, which means that the currents close locally across
field lines. 

\section{Conclusions}

We carried out numerical simulations of current sheet formation in a
quasi-separatrix layer using a simplified MHD model appropriate for
the quasi-static evolution of a low-$\beta$ plasma.
The setting under consideration has been investigated before
\cite{aulanier2005} and our results agree well with that work
as long as the current sheet structure is well resolved in both studies.
With the use of local adaptive mesh refinement (AMR), we were able to
follow the thinning of the current sheet further down to a scale which
is about one order of magnitude smaller than previously investigated.
In particular, our simulations reached a stage in which the maxima of
$|\vek{j}|$ in the upper part of the QSL grow significantly beyond the values
close to the photospheric boundaries, which gives clear evidence that
the most intense current densities actually can be expected
in the upper part of the QSL. 
This late stage is characterized by a relatively fast collapse of the
locally almost one-dimensional sheet with an approximately exponential
increase of $\text{max}(|\vek{j}|)$ in time, and
the evolution is no more quasi-static at this point.

When magnetic forces become significant  in this late stage of the current 
sheet formation, full nonlinear MHD dynamics will take place. 
Previous studies have addressed details of the local dynamics of such
current sheets using appropriate initial conditions and periodic systems
(e.g. \cite{grauerMarliani} and references therein).
There, one particular question has been whether the current density might
form a singularity in finite time, or whether its growth is limited to
merely e.g. exponential behavior.
On theoretical grounds, it could be shown that a dynamical alignment
between the velocity and the magnetic field would bound $|\vek{j}|$ to
exponential growth.
Analyzing our data further, we actually found indications of such an
alignment (not shown here), so that we expect to see a collapse of the
sheet with exponential growth, i.e. a continuation of the phase
observed between $t\approx 4.7$ and $\approx 5.5$ in
Fig.~\ref{figs:jmax_vs_t}, given that it could be resolved numerically.
At present however, even our AMR simulations are limited by the lack
of further resolution and by numerical side-effects like artificial
stabilizing diffusion.

The plasma flow pattern has been analyzed in a cut plane that is approximately
perpendicular to the local direction of the current density in the sheet:
If transformed into a frame that moves with the expanding structure,
it exhibits a shear pattern which arises from a vortex below and a
large-scale flow above.
The vortex maps to the vortical boundary driver, while
the large-scale flow is related to the slow overall expansion of the
magnetic structure.

In this paper, we have only addressed the case of one specific
boundary perturbation, namely a twisting motion at the footpoints.
We have also conducted a number of simulations with a
translational motion analogous to that used in \cite{aulanier2005},
and observed comparable behavior in these cases as well.
In particular, the achieved current densities continued to rise at similar rates
until they were restricted by finite grid resolution.

\begin{acknowledgments}
The authors would like to thank G. Aulanier for helpful discussions
during the preparation of the manuscript, and an anonymous referee for
comments that helped to improve it.
This work was supported by Deutsche Forschungsgemeinschaft through
Forschergruppe FOR 1048 and by the European Commission through the
Solaire network (MTRN-CT-2006-035484).
\end{acknowledgments}

%\nocite{*}


\begin{thebibliography}{14}
\expandafter\ifx\csname natexlab\endcsname\relax\def\natexlab#1{#1}\fi
\expandafter\ifx\csname bibnamefont\endcsname\relax
  \def\bibnamefont#1{#1}\fi
\expandafter\ifx\csname bibfnamefont\endcsname\relax
  \def\bibfnamefont#1{#1}\fi
\expandafter\ifx\csname citenamefont\endcsname\relax
  \def\citenamefont#1{#1}\fi
\expandafter\ifx\csname url\endcsname\relax
  \def\url#1{\texttt{#1}}\fi
\expandafter\ifx\csname urlprefix\endcsname\relax\def\urlprefix{URL }\fi
\providecommand{\bibinfo}[2]{#2}
\providecommand{\eprint}[2][]{\url{#2}}

\bibitem[{\citenamefont{{D{\'e}moulin}
  et~al.}(1996{\natexlab{a}})\citenamefont{{D{\'e}moulin}, {Henoux}, {Priest},
  and {Mandrini}}}]{demoulin1996a}
\bibinfo{author}{\bibfnamefont{P.}~\bibnamefont{{D{\'e}moulin}}},
  \bibinfo{author}{\bibfnamefont{J.~C.} \bibnamefont{{Henoux}}},
  \bibinfo{author}{\bibfnamefont{E.~R.} \bibnamefont{{Priest}}},
  \bibnamefont{and} \bibinfo{author}{\bibfnamefont{C.~H.}
  \bibnamefont{{Mandrini}}}, \bibinfo{journal}{Astronomy and Astrophysics}
  \textbf{\bibinfo{volume}{308}}, \bibinfo{pages}{643}
  (\bibinfo{year}{1996}{\natexlab{a}}).

\bibitem[{\citenamefont{{D{\'e}moulin}
  et~al.}(1996{\natexlab{b}})\citenamefont{{D{\'e}moulin}, {Priest}, and
  {Lonie}}}]{demoulin1996b}
\bibinfo{author}{\bibfnamefont{P.}~\bibnamefont{{D{\'e}moulin}}},
  \bibinfo{author}{\bibfnamefont{E.~R.} \bibnamefont{{Priest}}},
  \bibnamefont{and} \bibinfo{author}{\bibfnamefont{D.~P.}
  \bibnamefont{{Lonie}}}, \bibinfo{journal}{J. Geophs. Res.}
  \textbf{\bibinfo{volume}{101}}, \bibinfo{pages}{7631}
  (\bibinfo{year}{1996}{\natexlab{b}}).

\bibitem[{\citenamefont{{Milano} et~al.}(1999)\citenamefont{{Milano},
  {Dmitruk}, {Mandrini}, and {G\'omez}}}]{Milano1999}
\bibinfo{author}{\bibfnamefont{L.}~\bibnamefont{{Milano}}},
  \bibinfo{author}{\bibfnamefont{P.}~\bibnamefont{{Dmitruk}}},
  \bibinfo{author}{\bibfnamefont{C.}~\bibnamefont{{Mandrini}}},
  \bibnamefont{and}
  \bibinfo{author}{\bibfnamefont{D.}~\bibnamefont{{G\'omez}}},
  \bibinfo{journal}{Astrophys. J.} \textbf{\bibinfo{volume}{521}},
  \bibinfo{pages}{889} (\bibinfo{year}{1999}).

\bibitem[{\citenamefont{{Titov} and {Hornig}}(2002)}]{titovHornig2002}
\bibinfo{author}{\bibfnamefont{V.~S.} \bibnamefont{{Titov}}} \bibnamefont{and}
  \bibinfo{author}{\bibfnamefont{G.}~\bibnamefont{{Hornig}}},
  \bibinfo{journal}{Adv. Sp. Res.} \textbf{\bibinfo{volume}{29}},
  \bibinfo{pages}{1087} (\bibinfo{year}{2002}).

\bibitem[{\citenamefont{{Titov}}(2007)}]{titov2007}
\bibinfo{author}{\bibfnamefont{V.~S.} \bibnamefont{{Titov}}},
  \bibinfo{journal}{Astrophys. J.} \textbf{\bibinfo{volume}{660}},
  \bibinfo{pages}{863} (\bibinfo{year}{2007}), \eprint{arXiv:astro-ph/0703671}.

\bibitem[{\citenamefont{{Titov} et~al.}(2003)\citenamefont{{Titov},
  {Galsgaard}, and {Neukirch}}}]{Titov2003}
\bibinfo{author}{\bibfnamefont{V.~S.} \bibnamefont{{Titov}}},
  \bibinfo{author}{\bibfnamefont{K.}~\bibnamefont{{Galsgaard}}},
  \bibnamefont{and}
  \bibinfo{author}{\bibfnamefont{T.}~\bibnamefont{{Neukirch}}},
  \bibinfo{journal}{Astrophys. J.} \textbf{\bibinfo{volume}{582}},
  \bibinfo{pages}{1172} (\bibinfo{year}{2003}).

\bibitem[{\citenamefont{{Galsgaard} et~al.}(2003)\citenamefont{{Galsgaard},
  {Titov}, and {Neukirch}}}]{Galsgaard2003}
\bibinfo{author}{\bibfnamefont{K.}~\bibnamefont{{Galsgaard}}},
  \bibinfo{author}{\bibfnamefont{V.~S.} \bibnamefont{{Titov}}},
  \bibnamefont{and}
  \bibinfo{author}{\bibfnamefont{T.}~\bibnamefont{{Neukirch}}},
  \bibinfo{journal}{Astrophys. J.} \textbf{\bibinfo{volume}{595}},
  \bibinfo{pages}{506} (\bibinfo{year}{2003}).

\bibitem[{\citenamefont{Aulanier et~al.}(2005)\citenamefont{Aulanier, Pariat,
  and D{\'e}moulin}}]{aulanier2005}
\bibinfo{author}{\bibfnamefont{G.}~\bibnamefont{Aulanier}},
  \bibinfo{author}{\bibfnamefont{E.}~\bibnamefont{Pariat}}, \bibnamefont{and}
  \bibinfo{author}{\bibfnamefont{P.}~\bibnamefont{D{\'e}moulin}},
  \bibinfo{journal}{Astronomy and Astrophysics} \textbf{\bibinfo{volume}{444}},
  \bibinfo{pages}{961} (\bibinfo{year}{2005}).

\bibitem[{\citenamefont{{Aulanier} et~al.}(2006)\citenamefont{{Aulanier},
  {Pariat}, {D{\'e}moulin}, and {Devore}}}]{Aulanier2006}
\bibinfo{author}{\bibfnamefont{G.}~\bibnamefont{{Aulanier}}},
  \bibinfo{author}{\bibfnamefont{E.}~\bibnamefont{{Pariat}}},
  \bibinfo{author}{\bibfnamefont{P.}~\bibnamefont{{D{\'e}moulin}}},
  \bibnamefont{and} \bibinfo{author}{\bibfnamefont{C.~R.}
  \bibnamefont{{Devore}}}, \bibinfo{journal}{Sol. Phys}
  \textbf{\bibinfo{volume}{238}}, \bibinfo{pages}{347} (\bibinfo{year}{2006}).

\bibitem[{\citenamefont{Dreher and Grauer}(2005)}]{drehergrauer}
\bibinfo{author}{\bibfnamefont{J.}~\bibnamefont{Dreher}} \bibnamefont{and}
  \bibinfo{author}{\bibfnamefont{R.}~\bibnamefont{Grauer}},
  \bibinfo{journal}{Parallel Computing} \textbf{\bibinfo{volume}{31}},
  \bibinfo{pages}{913} (\bibinfo{year}{2005}).

\bibitem[{\citenamefont{{Arnold} et~al.}(2008)\citenamefont{{Arnold}, {Dreher},
  {Grauer}, {Soltwisch}, and {Stein}}}]{lukasFlarelab}
\bibinfo{author}{\bibfnamefont{L.}~\bibnamefont{{Arnold}}},
  \bibinfo{author}{\bibfnamefont{J.}~\bibnamefont{{Dreher}}},
  \bibinfo{author}{\bibfnamefont{R.}~\bibnamefont{{Grauer}}},
  \bibinfo{author}{\bibfnamefont{H.}~\bibnamefont{{Soltwisch}}},
  \bibnamefont{and} \bibinfo{author}{\bibfnamefont{H.}~\bibnamefont{{Stein}}},
  \bibinfo{journal}{Phys. Plasmas} \textbf{\bibinfo{volume}{15}},
  \bibinfo{pages}{042106} (\bibinfo{year}{2008}).

\bibitem[{\citenamefont{{Dedner} et~al.}(2002)\citenamefont{{Dedner}, {Kemm},
  {Kr{\"o}ner}, {Munz}, {Schnitzer}, and {Wesenberg}}}]{dedner2002}
\bibinfo{author}{\bibfnamefont{A.}~\bibnamefont{{Dedner}}},
  \bibinfo{author}{\bibfnamefont{F.}~\bibnamefont{{Kemm}}},
  \bibinfo{author}{\bibfnamefont{D.}~\bibnamefont{{Kr{\"o}ner}}},
  \bibinfo{author}{\bibfnamefont{C.}~\bibnamefont{{Munz}}},
  \bibinfo{author}{\bibfnamefont{T.}~\bibnamefont{{Schnitzer}}},
  \bibnamefont{and}
  \bibinfo{author}{\bibfnamefont{M.}~\bibnamefont{{Wesenberg}}},
  \bibinfo{journal}{Journal of Computational Physics}
  \textbf{\bibinfo{volume}{175}}, \bibinfo{pages}{645} (\bibinfo{year}{2002}).

\bibitem[{\citenamefont{{Shu} and {Osher}}(1988)}]{shuOsherRK}
\bibinfo{author}{\bibfnamefont{C.}~\bibnamefont{{Shu}}} \bibnamefont{and}
  \bibinfo{author}{\bibfnamefont{S.}~\bibnamefont{{Osher}}},
  \bibinfo{journal}{J. Comp. Phys.} \textbf{\bibinfo{volume}{77}},
  \bibinfo{pages}{439} (\bibinfo{year}{1988}).

\bibitem[{\citenamefont{{Grauer} and {Marliani}}(2000)}]{grauerMarliani}
\bibinfo{author}{\bibfnamefont{R.}~\bibnamefont{{Grauer}}} \bibnamefont{and}
  \bibinfo{author}{\bibfnamefont{C.}~\bibnamefont{{Marliani}}},
  \bibinfo{journal}{Phys. Rev. Lett.} \textbf{\bibinfo{volume}{84}},
  \bibinfo{pages}{4850} (\bibinfo{year}{2000}).

\end{thebibliography}
\end{document}